\documentclass[10pt,journal]{IEEEtran}
\IEEEoverridecommandlockouts
\usepackage{cite}
\usepackage{amsmath,amssymb,amsfonts}
\usepackage{algorithmic}
\usepackage{float}
\usepackage{graphicx}
\usepackage{textcomp}
\usepackage{xcolor}
\usepackage{stfloats}
\usepackage[linesnumbered,ruled,vlined]{algorithm2e}
\usepackage{array}
\usepackage{multirow}
\usepackage{caption}
\usepackage{flushend}
\usepackage{mathrsfs}
\usepackage{verbatim}
\usepackage{booktabs}
\usepackage{tabularx}
\usepackage{array}
\usepackage{makecell}

\def\BibTeX{{\rm B\kern-.05em{\sc i\kern-.025em b}\kern-.08em
    T\kern-.1667em\lower.7ex\hbox{E}\kern-.125emX}}

\def\BibTeX{{\rm B\kern-.05em{\sc i\kern-.025em b}\kern-.08em
    T\kern-.1667em\lower.7ex\hbox{E}\kern-.125emX}}

\makeatletter

\newcommand{\Rmnum}[1]{\expandafter\@slowromancap\romannumeral #1@}
\makeatother

\begin{document}
\title{Fluid Antenna Systems for ISAC Design: A Survey}

\author{Yueling~Zhao,
        Ping~Yang,~\IEEEmembership{Senior Member, IEEE,}
        Hong~Niu,
        Shuaixin~Yang,
        Yue~Xiao,~\IEEEmembership{Member, IEEE,}    Gang~Wu,~\IEEEmembership{Senior~Member, IEEE,}
        M\'{e}rouane~Debbah,~\IEEEmembership{Fellow, IEEE,}
        and Ertugrul~Basar,~\IEEEmembership{Fellow, IEEE}
\vspace{-3 em}

\thanks{Y. Zhao, P. Yang (corresponding author), S. Yang, Y. Xiao and G. Wu are with the National Key Laboratory of Wireless Communications, University of Electronic Science and Technology of China, Chengdu 611731, China .

H. Niu (corresponding author) is with the School of Electrical and Electronic Engineering, Nanyang Technological University, Singapore 639798.

M. Debbah is with the Research Institute for Digital Future, Khalifa University, 127788 Abu Dhabi, United Arab Emirates;

E. Basar is with the Tampere Wireless Research Centre, Department of Electrical Engineering, Tampere University, 33720 Tampere, Finland, also holds an adjunct position with the Department of Electrical and Electronics Engineering, Ko\c{c} University, Sariyer, 34450 Istanbul, T\"{u}rkiye. 

\textcopyright{} 2026 IEEE. Personal use of this material is permitted. Permission from IEEE must be obtained for all other uses.}
}
\vspace{-2em}
%\vspace{-10mm}
\maketitle

\begin{abstract}
Fluid antenna systems (FAS) introduce dynamic positional degrees of freedom (DoFs), extending the spatial domain from a static and discrete resource to a quasi-continuous and reconfigurable design space. This spatial flexibility provides an effective means of alleviating the geometric rigidity and functional conflicts that constrain integrated sensing and communication (ISAC) systems. Against this backdrop, we provide a comprehensive overview of FAS-enabled ISAC, tracing the evolution from theoretical performance limits to practical transceiver architectures. To illustrate the surveyed design evolution, we further propose the fluid antenna multi-mode modulation (FA-MM) framework as a representative case study of multidimensional physical-layer design for FAS-ISAC. Finally, several open challenges are discussed, with emphasis on predictive sensing, adaptive coordination, and secure spatial-state management in future FAS-ISAC systems.
\end{abstract}


\vspace{-0.5em}
\section{Introduction}

\IEEEPARstart{A}{s} a key usage scenario for sixth-generation (6G) networks, integrated sensing and communication (ISAC) extends wireless networks from communication-centric platforms to dual-functional systems capable of both data transmission and sensing~\cite{ISAC}. While spectrum and hardware sharing enhance efficiency, they introduce a fundamental tension: communication favors stable, high-gain links for data capacity, whereas radar sensing requires agile beams for target tracking. Addressing this challenge demands additional degrees of freedom (DoFs) beyond conventional time- and frequency-domain resources, emphasizing the critical role of the spatial domain in balancing communication and sensing objectives.

Spatial-domain DoFs have been a primary driver of wireless performance over the past decade, evolving from conventional multiple-input multiple-output (MIMO) systems to extremely large-scale MIMO (XL-MIMO) architectures. These advances enhance spectral efficiency (SE) and network capacity but come at the cost of increased hardware complexity, power consumption, and physical footprint. For ISAC systems, the fixed geometry of traditional fixed-position antennas (FPAs) limits the effective exploitation of spatial DoFs, constraining the trade-off between communication capacity and sensing resolution. Such limitations underscore the need for architectures that enable flexible spatial-state management, setting the stage for dynamically reconfigurable solutions~\cite{Tuo,FAS}.

Fluid antenna systems (FAS) address these constraints by introducing dynamic positional flexibility~\cite{FAS}. Such architectures encompass reconfigurable antennas capable of adjusting radiating positions, physical shape, or radiation characteristics, with implementations ranging from conductive fluids to pixel-based reconfigurable antennas (PRAs)~\cite{Tuo}. This adaptability provides three key capabilities: spatial liberation, enabling quasi-continuous positioning to bypass discrete grid constraints; resource coordination, allowing dynamic allocation of antenna states to mitigate interference and dual-functional trade-offs; and intelligent adaptation, supporting integration with emerging 6G scenarios. In particular, FAS can exploit near-field focusing for enhanced spatial resolution, while enabling artificial intelligence (AI)-driven optimization for high-dimensional port selection with reduced online complexity~\cite{FAS}.

Recent studies highlight that FAS, when applied to ISAC, can leverage positional DoFs to alleviate resource contention and coordinate coexisting tasks~\cite{Yue}. By expanding the spatial design space, FAS allow antenna configurations to be dynamically adapted, reconciling the distinct spatial requirements of sensing and communication while mitigating the geometric conflicts of conventional FPAs~\cite{Tuo}. More fundamentally, FAS allow the spatial geometry itself to become adaptive, enabling ISAC systems to jointly optimize signal transmission and spatial configurations. Under this paradigm, antenna placement, port selection, and beamforming are jointly orchestrated to enhance the performance of multifunctional networks~\cite{Yue}.

Motivated by these advantages, this article provides a comprehensive overview of the role of FAS in ISAC networks. Focusing on FAS-enabled ISAC, this survey organizes recent advances ranging from theoretical foundations to practical system implementations. We review representative architectures, enabling techniques, and optimization methodologies that exploit positional DoFs for dual-functional performance enhancement. To further bridge the surveyed concepts and practical physical-layer design, we propose the fluid antenna multi-mode modulation (FA-MM) framework, illustrating how positional states and signal mappings can be jointly organized within a unified design. Finally, we discuss open challenges and future directions, highlighting how future FAS-ISAC systems may evolve from static spatial optimization toward adaptive, coordinated, and secure spatial-state management.
\vspace{-0.5em}
\section{Rethinking Spatial DoFs in ISAC: The Fluid Antenna Paradigm}
\vspace{-0.1em}
The physical-layer realization of ISAC relies heavily on the efficient use of spatial DoFs. Conventional arrays provide these DoFs through fixed geometries, making it difficult to adapt the spatial response to concurrent sensing and communication demands. This section first reviews how FAS relax these geometric constraints through dense, quasi-continuous spatial reconfiguration within a confined aperture, and then examines the resulting implications for ISAC, including compact spatial-DoF expansion, hardware-efficient interference suppression, and improved dual-functional reliability.
\vspace{-1em}
\subsection{The Spatial Dilemma: FPA Constraints in ISAC}

By integrating radar sensing and communication within shared spectrum, hardware, and waveform resources, ISAC improves SE and enables dual-functional coordination~\cite{ISAC}. Despite these advantages, the concurrent use of shared time-frequency resources forces ISAC systems to rely heavily on the spatial domain for signal separation. From an array-processing perspective, sensing and communication impose distinct radiation requirements: target sensing requires focused beams with low sidelobes for precise angle estimation and clutter suppression, whereas multi-user communication prioritizes spatial multiplexing and stable array gains. Consequently, synthesizing sensing beams while forming spatial nulls for interference mitigation places a significant demand on the finite spatial DoFs of conventional arrays, limiting joint sensing-communication performance in dense scenarios.

Conventionally, alleviating spatial-DoF saturation requires increasing the antenna count or enlarging the aperture. For FPA-based arrays, however, fixed element locations and half-wavelength-scale spacing constraints tie additional spatial DoFs to larger physical footprints, making such scaling difficult for compact platforms. Once deployed, their discrete geometry fixes the spatial samples and observation manifold, limiting the ability to refine sensing resolution, reshape interference patterns, and balance communication-sensing requirements. Static arrays may also be susceptible to localized blockages and deep fades, as their fixed radiating positions limit adaptation to favorable propagation conditions. These geometric rigidities motivate alternative architectures that can enrich spatial states without relying solely on macroscopic aperture expansion.
\vspace{-1em}
\subsection{FAS: Spatial Liberation and DoF Expansion}
To address the discrete and static spatial constraints of conventional arrays, FAS have emerged as a reconfigurable physical-layer architecture~\cite{FAS}. Enabled by advancements in conductive liquid metals (e.g., eutectic gallium-indium) and reconfigurable radio frequency (RF) pixel grids, FAS permit the fine-grained relocation or state reconfiguration of the active antenna port within a confined physical aperture~\cite{Tuo}. This mechanism extends array operation from a finite set of stationary elements to a dense, quasi-continuous spatial domain. By selecting physical ports or electronically emulating radiating positions at sub-wavelength scales, FAS can reduce the dependence on fixed antenna locations in conventional FPAs. Rather than relying on macroscopic aperture expansion, this sub-wavelength agility provides the physical basis for dense spatial-state reconfiguration within a compact aperture. In this sense, FAS shift array geometry from a fixed hardware constraint to an optimization resource.

Exploiting this expanded spatial-state space, FAS provide an effective means of accommodating the distinct radiation requirements of radar and communication tasks. For target sensing, ultra-dense spatial sampling enables the system to capture fine-grained phase variations across the physical aperture. Such dense observations can improve parameter estimation accuracy relative to standard discrete grids. From a communication perspective, FAS utilize this high-resolution state space to enhance spatial selectivity. By activating ports at channel-favorable locations, FAS can exploit sub-wavelength channel variations to suppress multi-user interference while enhancing desired-signal reception. Taken together, these properties provide FAS with finer spatial controllability than conventional FPAs. In practical implementations, these gains are conditioned by finite-aperture correlation, mutual coupling, switching losses, and the overhead of position-dependent CSI acquisition and calibration.

Beyond sensing and communication controllability, the spatial agility of FAS also helps counter environmental impairments in dual-functional operations. Localized deep fades and blockages may simultaneously degrade communication reliability and sensing continuity. By leveraging micro-scale port mobility, FAS introduce dynamic spatial adaptability at the physical layer. Given that multipath fading fluctuates significantly over fractional wavelengths, a sub-wavelength displacement of the active radiating position may allow the system to escape deep fades. This mechanism improves communication robustness and sensing reliability even in complex scattering environments. Ultimately, by transitioning the spatial domain from a strictly discrete grid into a quasi-continuous state, FAS alleviate the spatial rigidities inherent to conventional FPAs.

\vspace{-1.5em}
\subsection{Towards FAS-ISAC: Integration and Benefits}

When integrated into ISAC, FAS translate the positional flexibility discussed above into a compact mechanism for coordinating sensing and communication resources. In contrast to conventional FPA-based systems, which often rely on larger apertures or additional RF hardware to support target sensing and multi-user communication simultaneously, FAS exploit physical-layer reconfigurability within a confined aperture to relieve spatial resource contention. In this sense, spatial configurations become adaptive optimization variables rather than fixed hardware constraints, giving rise to several benefits for FAS-ISAC:

\subsubsection{Decoupling Spatial Capabilities from Physical Constraints} FAS reduce the dependence of ISAC spatial capability on physical footprint. In conventional FPA systems, high-fidelity sensing and low-correlation communication often rely on larger apertures or wider antenna separation. By enabling dense spatial-state reconfiguration within a compact aperture, FAS support spatial diversity for both tasks without proportional hardware scaling. This flexibility enables footprint-constrained platforms, such as unmanned aerial vehicles (UAVs) and vehicular terminals, to support high-precision sensing and robust communication with compact antenna structures.

\subsubsection{Hardware-Efficient Spatial Nulling for Sensing Preservation} FAS help shift part of the interference-suppression burden from digital beamforming to spatial-domain adaptation. In conventional systems, suppressing communication interference typically consumes spatial DoFs and RF chains, reducing the resources available for radar sensing. Through active-port repositioning, FAS can suppress interfering signals within the propagation environment. This reduces the need for digital nulling, allowing more RF hardware and processing resources to be preserved for continuous environmental sensing without compromising radar performance.

\subsubsection{Mitigating Dual-Functional Outages} FAS introduce a physical-layer evasion mechanism to improve dual-functional reliability under environmental uncertainties. In complex scattering environments, localized deep fades and blockages may simultaneously disrupt communication transmission and sensing echoes, with radar sensing being particularly sensitive due to two-way path loss. By exploiting sub-wavelength spatial evasion~\cite{FAS}, FAS allow the active port to bypass unfavorable propagation regions through micro-scale displacement. This adaptability helps prevent localized impairments from causing dual-functional interruptions, sustaining target tracking and data transmission under blockage and dynamic fading conditions.

Taken together, these benefits indicate that FAS extend spatial reconfigurability from a physical-layer feature to a system design mechanism for ISAC. Instead of remaining passively constrained by array geometry and hardware scaling, spatial configurations become adaptive resources that can be dynamically coordinated to support the diverse and often conflicting requirements of sensing and communication.
\vspace{-1em}
\section{Advancing FAS-ISAC: Theoretical, Algorithmic, and Architectural Evolutions}

Building on the preceding discussion of the benefits of FAS-based systems, this section traces the technical evolution of FAS-ISAC. As illustrated in Fig. 1, the review begins with fundamental performance limits, followed by AI-driven spatial optimization, emerging scenarios and cross-domain synergies, and finally transceiver architectures and hardware implementations. This progression illustrates how positional flexibility evolves from a reconfigurable spatial capability into a practical system design for ISAC.

\captionsetup[figure]{name={Fig.},labelsep=period}
\begin{figure*}[t]
%\centerline{\includegraphics[width = 7.1in]{Evolution}}
\centerline{\includegraphics[width = 6in]{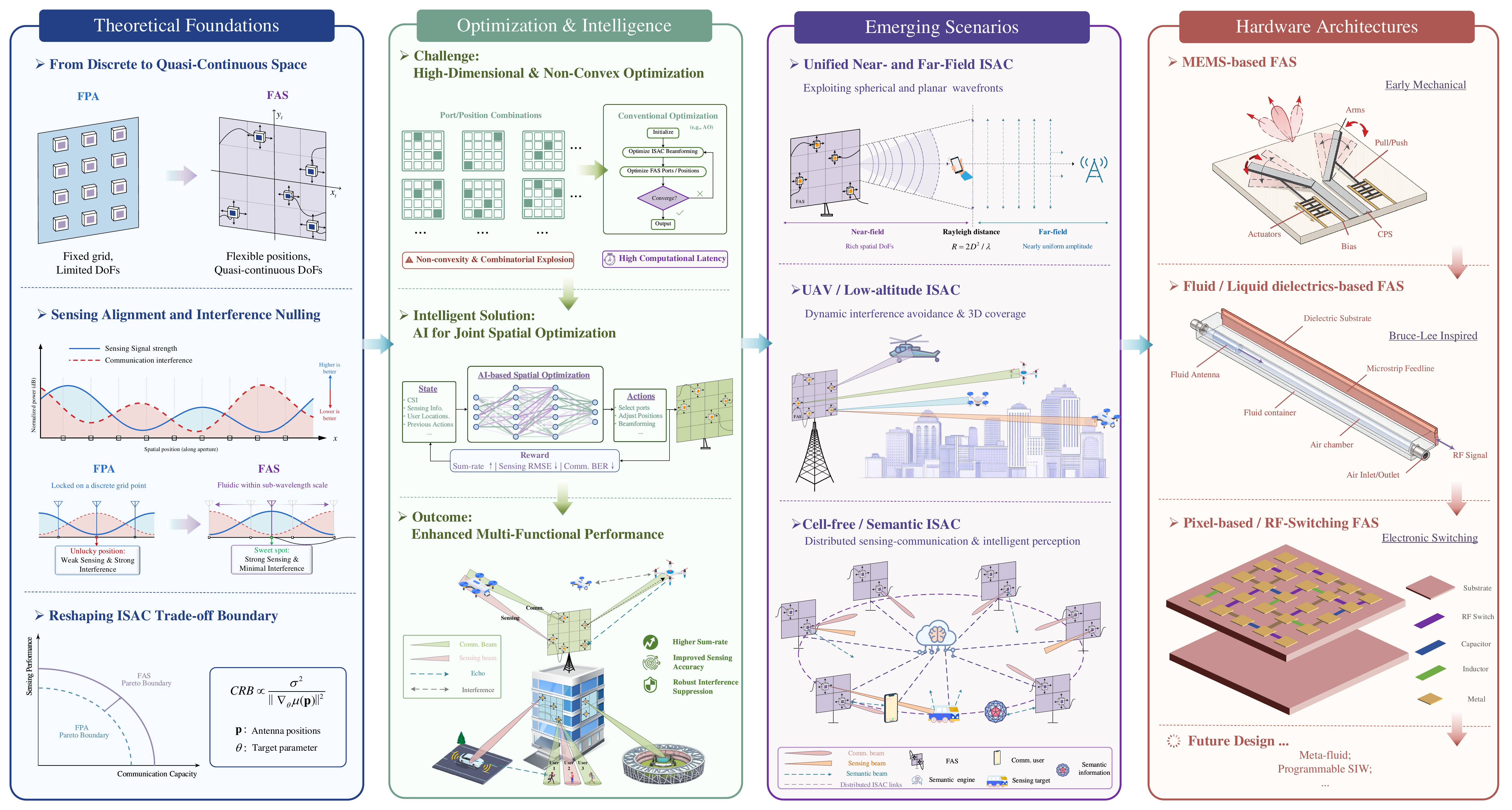}}
\caption{{\small{Evolution of FAS-ISAC: From theoretical foundations to hardware implementation.}}}
\label{fig}
\vspace{-2em}
\end{figure*}
%\vspace{-20mm}

\vspace{-1em}
\subsection{Fundamental Limits: Trade-off, Estimation, and Robustness}
Recent FAS-ISAC studies have characterized how positional flexibility reshapes the achievable communication-sensing performance region. In conventional FPAs, fixed antenna locations constrain spatial-DoF allocation between target sensing and communication transmission. Dense spatial-state reconfiguration mitigates this limitation by allowing antenna positions and ports to participate as adaptive variables in balancing dual-functional requirements. This subsection reviews theoretical studies on trade-off shifting, sensing bounds, and robustness under channel uncertainty.

\subsubsection{Communication-Sensing Trade-off Boundaries}
Representative studies first quantify the expansion of the communication-sensing Pareto boundary enabled by continuous spatial mobility. The authors in \cite{Trade-Off} suggest that port selection in FAS can alter how spatial resources are shared between interference suppression and target sensing. By formulating a joint optimization of transmit beamforming and antenna positioning, this study indicates that sub-wavelength spatial agility can mitigate spatial competition between multi-user interference suppression and target sensing. This capability shifts the ISAC trade-off, potentially expanding the achievable Pareto frontier compared to conventional FPAs.  

Building on this trade-off perspective, the work presented in \cite{Zhou_FA_ISAC} investigates the maximization of communication sum-rate under predefined sensing beam gain constraints for FAS-assisted downlinks. To resolve the non-convex coupling between continuous antenna coordinates and digital beamforming vectors, an alternating optimization (AO) framework leveraging successive convex approximation (SCA) is employed. The results indicate that exploiting the continuous spatial configuration space can provide measurable gains in sum-rate performance while satisfying                           sensing requirements. This analysis highlights how positional DoF optimization can coordinate dual-functional demands while satisfying sensing constraints.

\subsubsection{Sensing Accuracy and Clutter Suppression}
Beyond overall system capacity, FAS provide substantial improvements in the precision of target parameter estimation. The authors of \cite{Ye_SCNR} explore the fundamental limits of clutter suppression by maximizing the sensing-to-clutter-plus-noise ratio (SCNR) in MIMO-ISAC systems. Their analysis demonstrates that continuous spatial repositioning of antennas enables deep spatial nulling against dynamic environmental clutter, isolating the target echo without significantly compromising the concurrent communication signal-to-interference-plus-noise ratio (SINR). 

Extending the theoretical bounds of estimation accuracy,~\cite{CRB} derives the Cramér-Rao bound (CRB) for direction-of-arrival estimation in FAS-empowered uplink ISAC networks. By treating continuous antenna coordinates as optimization variables, this mathematical framework minimizes the CRB. It suggests that the high-resolution spatial diversity enabled by FAS can overcome the resolution limitations of static uniform linear arrays, thus improving the attainable estimation accuracy under suitable spatial configurations.

\subsubsection{Robustness Under Channel Uncertainty} 
While the aforementioned theoretical bounds assume ideal channel conditions, the practical deployment of FAS-ISAC requires robustness against channel estimation errors under continuous spatial reconfiguration. Addressing this, research in \cite{Hao_Imperfect_CSI} investigates the joint optimization of antenna positioning and dual-functional beamforming under both perfect and imperfect channel state information (CSI). By formulating a robust optimization framework, this study demonstrates that the spatial agility of FAS can compensate for CSI uncertainties. This joint adaptation helps maintain communication link reliability and sensing beam accuracy, even when exact channel matrices are degraded. Consequently, this robust design shows that the capacity and estimation-accuracy gains associated with FAS can be sustained in non-ideal propagation environments.

The key theoretical advantages demonstrated by these foundational FAS-ISAC frameworks include:
\begin{itemize}
  \item Dense positional DoFs expand the communication-sensing trade-off boundary without proportional hardware scaling.
  \item Sub-wavelength agility improves clutter suppression and reduces the CRB, thereby enhancing sensing accuracy.
  \item Spatial diversity provides physical-layer robustness against imperfect CSI and channel uncertainty.
\end{itemize}

Collectively, these studies establish the theoretical foundations of FAS-ISAC by characterizing the role of spatial reconfigurability in communication-sensing trade-offs, sensing accuracy, and robustness. These insights motivate subsequent algorithmic and implementation studies.

% Compact two-entry cell
\newcommand{\twocell}[2]{%
  \parbox[t]{\linewidth}{%
    #1\par\vspace{-0.25ex}#2%
  }%
}

% Gap between the double-column table and the following text
\setlength{\dbltextfloatsep}{15pt plus 1pt minus 1pt}

% ---------- Table ----------
\begin{table*}[!t]
\centering
\caption{Taxonomy and Critical Comparison of Representative FAS-ISAC Studies}
\label{tab:fas_isac_taxonomy}

\fontsize{6.6}{7.0}\selectfont
\renewcommand{\arraystretch}{0.88}
\setlength{\tabcolsep}{1.25pt}

\begin{tabular}{
>{\raggedright\arraybackslash}p{0.115\textwidth}
>{\raggedright\arraybackslash}p{0.225\textwidth}
>{\raggedright\arraybackslash}p{0.210\textwidth}
>{\raggedright\arraybackslash}p{0.245\textwidth}
>{\raggedright\arraybackslash}p{0.155\textwidth}
}
\toprule
\textbf{Taxonomy (Refs.)} &
\textbf{System / FAS} &
\textbf{C/S Objective} &
\textbf{Variables / Method} &
\textbf{Assumptions / Limits} \\
\midrule

\textbf{Comm.-oriented} [5], [6] &
\twocell
{[5] MU DL; BS Tx ports}
{[6] SU DL; BS/UE FAs} &
\twocell
{[5] Power min.; C/S SNR constraints}
{[6] Rate max.; sensing gain} &
\twocell
{[5] Tx BF + port selection; AO}
{[6] Tx BF + FA positions; AO} &
Perfect CSI; ideal reconfiguration
\\
\midrule

\textbf{Rx-FAS sensing} [7], [8] &
\twocell
{[7] MU DL; sensing Rx-FAs}
{[8] MU UL; BS Rx-FAs} &
\twocell
{[7] SCNR max.; user SINR}
{[8] CRB min.; UL constraints} &
\twocell
{[7] Tx BF + Rx-FA positions; AO}
{[8] Power + Rx BF + FA positions; PDD/MM} &
Perfect CSI; known target response; ideal positioning
\\
\midrule

\textbf{Robust design} [9] &
MU DL; BS Tx-FAs + fixed Rx &
Sensing-SNR max.; robust SINR &
Dual BF + FA positions; AO/SCA &
Bounded CSI; ideal motion
\\
\midrule

\textbf{AI-driven} [10], [11] &
\twocell
{[10] MU DL; multi-port BS FAS}
{[11] MU multitarget; BS FAs} &
\twocell
{[10] Sum-rate max.; sensing constraints}
{[11] C/S utility max.} &
\twocell
{[10] Port selection + BF; DRL}
{[11] BF + FA positions; BCD/DDPG} &
Perfect/partial CSI; training cost
\\
\midrule

\textbf{Emerging scenarios} [12], [13] &
\twocell
{[12] Near-field ISAC; BS FAs}
{[13] Multi-BS LAE; UAV targets} &
\twocell
{[12] Data utility max.; joint constraints}
{[13] Weighted C/S rate max.} &
\twocell
{[12] BF + FA + semantic vars.; AO/SCA}
{[13] C/S BF + FA positions; heuristic} &
Geometry-based CSI; mobility/complexity limits
\\
\midrule

\textbf{Hardware} [14] &
MU ISAC; meta-fluid BS &
Reliability; inverse scattering &
Activation + reconstruction &
PIN-diode SIW; single RF
\\
\bottomrule
\end{tabular}

\vspace{0.35mm}

\parbox{0.985\textwidth}{%
\fontsize{5.05}{5.40}\selectfont
\textit{Abbreviations:}
AO, alternating optimization;
BF, beamforming;
BS/UE, base station/user equipment;
C/S, communication/sensing;
DL/UL, downlink/uplink;
FA, fluid antenna;
MU/SU, multi-user/single-user;
Tx/Rx, transmit/receive;
BCD, block coordinate descent;
DDPG, deep deterministic policy gradient;
DRL, deep reinforcement learning;
MM, majorization--minimization;
PDD, penalty dual decomposition;
SCA, successive convex approximation;
SCNR, sensing-to-clutter-plus-noise ratio;
SNR, signal-to-noise ratio.
}
\vspace{-4.5mm}
\end{table*}

\vspace{-1.5em}
\subsection{Algorithmic Design: AI-Driven Spatial Optimization}

While the theoretical bounds discussed previously establish the performance limits of FAS-ISAC, practical exploitation of these benefits requires efficient algorithms to identify favorable antenna configurations in high-dimensional spatial search spaces. Conventional mathematical tools, such as AO, incur computational overhead that scales rapidly with the number of antennas and users. This latency challenges real-time dual-functional operation, motivating AI-assisted spatial decision-making that shifts part of the optimization burden from online iteration to learned policy inference.

The study in \cite{USR_DRL} addresses this complexity using a deep reinforcement learning (DRL) framework for FAS-assisted multi-user MIMO-ISAC networks. By reformulating the joint optimization of fluid antenna port selection and transmit beamforming as a Markov decision process, the DRL agent learns to dynamically allocate spatial resources. This approach alleviates the computational burden of high-dimensional spatial search by learning a data-driven alternative to iterative solvers, enabling efficient sum-rate improvement under sensing constraints.

Further advancing this intelligent positioning strategy, research in \cite{IAP_DRL} investigates DRL-enabled dynamic antenna repositioning tailored for FAS-aided ISAC environments. This algorithmic design maps dynamic CSI and environmental sensing requirements to favorable fluid antenna coordinates. The results indicate that learning-based models can approach high-quality configurations, efficiently maintaining multi-user interference suppression and target tracking capabilities by exploiting continuous spatial reconfigurability.

The integration of AI into FAS-ISAC provides specific algorithmic advantages:
\begin{itemize}
    \item DRL models reduce the reliance on latency-prone iterative methods, enabling faster spatial port selection for dynamic ISAC scenarios.
    \item Learning-based algorithms handle the non-convex coupling between spatial configuration and digital beamforming.
\end{itemize}

These studies suggest that AI-driven optimization offers an effective alternative to conventional iterative methods for real-time spatial adaptation in FAS-ISAC. Future developments will require integrating learning-based optimization with practical deployment constraints.
\vspace{-1em}
\subsection{Emerging Scenarios and Cross-Domain Synergies}

The integration of ISAC into advanced 6G scenarios introduces complex propagation environments that challenge conventional FPAs. The positional flexibility of FAS provides an adaptive spatial mechanism for handling dynamic and localized propagation conditions. This subsection reviews representative studies that extend FAS-ISAC from conventional wireless systems to emerging scenarios, including near-field propagation and the low-altitude economy (LAE).

In near-field communications, the conventional plane-wave assumption becomes invalid, and spherical wavefronts make antenna position closely coupled with angle-range channel perception. In this setting, the study in \cite{NFC} investigates an architecture that integrates sensing, computing, and semantic communication using FAS. Specifically, the considered framework combines digital beamforming, computational resource allocation, and FAS port selection. By exploiting the distance-dependent characteristics of spherical wavefronts, FAS enable spatial repositioning to actively capture localized channel variations in both angle and range. This approach reduces latency constraints in cross-tier data processing, improving target sensing accuracy while supporting semantic-aware processing for intelligent applications. These results highlight the potential of FAS to exploit distance-dependent propagation characteristics for integrated communication, sensing, and computing.

Beyond near-field channels, FAS also present potential for mitigating air-to-ground interference in mobile, three-dimensional deployments, such as low-altitude networks. Aerial platforms are often subject to line-of-sight interference, which degrades both communication links and radar echo detection. Research in \cite{Zuo_UAV} proposes a joint design for dual-functional beamforming and FAS port switching in aerial ISAC nodes. This algorithmic approach demonstrates that dynamic spatial repositioning provides additional spatial DoFs for reshaping interference geometry, enabling sensing beam steering and spatial null formation towards unintended ground users. As a result, this method suppresses interference, supporting more robust communication and safer operation for low-altitude networks. This study illustrates how spatial reconfigurability can accommodate the dynamic propagation conditions of aerial ISAC networks.

The deployment of FAS-ISAC in these emerging scenarios provides key cross-domain benefits:
\begin{itemize}
  \item By exploiting spherical wavefronts, FAS enable localized spatial sensing in both angle and range, improving near-field ISAC capability.
  \item Dynamic spatial repositioning allows FAS to reshape interference patterns in mobile three-dimensional deployments, improving dual-functional reliability in low-altitude networks.
\end{itemize}

Together, these studies demonstrate that FAS-ISAC extends naturally to emerging scenarios with distinct propagation characteristics, broadening its application scope.

\vspace{-1.25em}
\subsection{Transceiver Architectures and Hardware Implementations}

Realizing theoretical advancements in practical ISAC systems requires viable physical implementations. Many early FAS models rely on liquid-metal displacement or mechanical actuation. However, these physical approaches face non-negligible latency, environmental sensitivity, and hardware wear. Consequently, recent research has increasingly shifted toward pixelated architectures that realize spatial reconfigurability through rapid electronic switching, moving FAS toward electronically programmable spatial platforms.

An implementation in \cite{Meta_Fluid} utilizes positive-intrinsic-negative (PIN) diodes integrated into a substrate-integrated waveguide (SIW) structure. By selectively activating elements, this meta-fluid architecture realizes fluid-like spatial reconfigurability using a single RF chain. The study further develops a single-RF multi-pattern activation scheme for interference management and an inverse scattering algorithm for high-resolution sensing. Full-wave simulations validate robust multi-user ISAC operation enabled by this electronically reconfigurable design.

Key insights from this hardware-efficient approach include:
\begin{itemize}
  \item Electronic switching across discrete ports efficiently emulates spatial reconfiguration while reducing the latency associated with mechanically actuated designs.
  \item The co-design of programmable SIW structures and multi-pattern activation algorithms provides a scalable single-RF architecture for integrated sensing and communication.
\end{itemize}

Table~I provides a structured taxonomy and comparison of representative FAS-ISAC studies. The comparison reveals an evolution from communication- and sensing-oriented optimization~\cite{Trade-Off,Zhou_FA_ISAC,Ye_SCNR,CRB,Hao_Imperfect_CSI}, to learning-assisted spatial optimization~\cite{USR_DRL,IAP_DRL}, and further to emerging scenarios and hardware implementations~\cite{NFC,Zuo_UAV,Meta_Fluid}. Together, these directions translate spatial reconfigurability from theoretical gains into practical algorithms, application designs, and hardware architectures. Across these directions, most existing designs treat spatial reconfiguration as an optimization variable for beamforming, port selection, or antenna positioning. Its integration with signal-domain modulation, however, has received less attention in existing FAS-ISAC studies. Section~IV presents a representative case study of this design perspective.
\vspace{-1em}
\section{Representative Case Study: FA-MM for Joint DoF Exploitation in FAS-ISAC}

Building on the preceding comparison between FPAs and FAS, an important design consideration for FAS-ISAC is how antenna reconfigurability can be incorporated into physical-layer design for joint communication and sensing. Existing FAS modulation schemes have demonstrated the potential of exploiting antenna-position indices, while broader opportunities lie in the enlarged physical-layer state space enabled by multidimensional state design. FA-MM is adopted in this section as a representative case study, illustrating the joint use of spatial reconfigurability and signal-domain modulation for dual-functional design.

\captionsetup[figure]{name={Fig.},labelsep=period}
\begin{figure*}[t]
%\centerline{\includegraphics[width = 6.6 in]{FA-MM.pdf}}
\centerline{\includegraphics[width = 5 in]{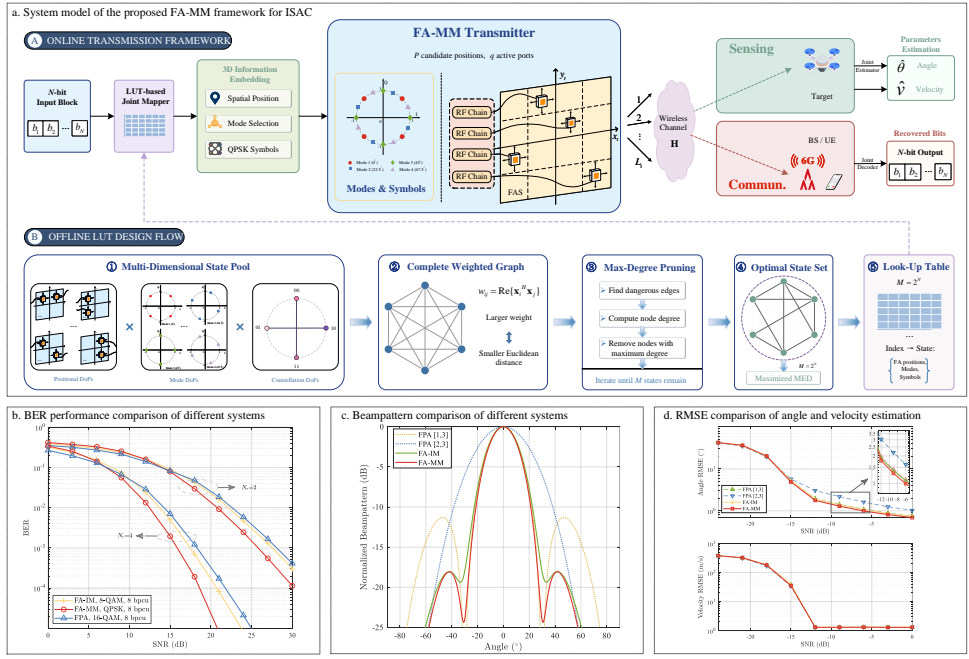}}
\caption{{\small{The proposed FA-MM framework: (a) system model, (b) BER performance comparison with conventional schemes, (c) normalized beampattern comparison, and (d) sensing RMSE performance for angle and velocity estimation.}}}
\label{fig}
\vspace{-1.5em}
\end{figure*}
\vspace{-1.25em}

\subsection{From FPAs to Joint DoF Exploitation in FAS-ISAC}

As discussed in Section III-D, exploiting the reconfigurability of FAS calls for modulation strategies that couple spatial states with signal-domain mappings. Index modulation~(IM)-based FAS architectures, represented by FA-IM, take a step in this direction by embedding antenna-position indices into conventional symbol mapping~\cite{Zhu}. However, antenna positions in such designs primarily serve as discrete modulation indices, while their joint use with other physical-layer DoFs for ISAC has received comparatively less attention. FA-MM provides a representative realization of this design perspective by jointly organizing positional, mode, and constellation DoFs within a unified modulation framework.
\vspace{-1.25em}

\subsection{FA-MM: Multidimensional DoF Integration}

As shown in Fig.~2(a), FA-MM jointly integrates positional, mode, and constellation DoFs into a multidimensional candidate state space. These dimensions are then organized into a unified look-up table (LUT) for modulation.

To construct the LUT, the candidate states are represented as a weighted graph, where edge weights characterize the similarity between multidimensional states. Graph-based pruning is then employed to remove highly correlated candidate states while preserving distinguishable ones, resulting in a unified mapping among antenna positions, transmission modes, and constellation symbols. This offline pruning constitutes the main LUT-construction overhead and avoids exhaustive state search during online transmission.

The resulting LUT increases state separability while reducing spatial correlations, providing a unified framework for jointly supporting communication and sensing within a common physical-layer design.
\vspace{-1.25em}
\subsection{Representative Performance Evaluation}
\vspace{-0.25em}

The representative communication and sensing performance of FA-MM is evaluated against conventional FPA systems and FA-IM under an equivalent SE of 8 bpcu. All schemes use the same linear FAS aperture spanning 1.5 wavelengths, with four candidate ports and two active RF chains, under identical hardware and SE constraints. Communication performance is evaluated over i.i.d. Rayleigh fading channels, whereas sensing considers deterministic target echoes corrupted by additive white Gaussian noise. All communication codebooks are normalized to the same average transmit energy, with the SNR varied through the noise variance. At the receiver, FA-MM and FA-IM employ joint maximum-likelihood detection over their respective codebooks for BER calculation.

Fig.~2(b) shows that FA-MM achieves an approximately 2~dB gain over the FPA baseline at a BER of $10^{-3}$ with two receive antennas. With four receive antennas, the gain reaches approximately 4.5~dB over FPA and 1.5~dB over FA-IM at a BER of $10^{-5}$. Figure~2(c) shows sharper target-directed beampatterns with lower sidelobes than both FPA configurations and FA-IM. Consistent improvements are also observed in sensing, where Fig.~2(d) demonstrates the lowest angle-estimation RMSE while maintaining comparable velocity-estimation accuracy. These results demonstrate that multidimensional state design can simultaneously improve communication reliability and sensing performance under identical hardware and SE constraints.

This representative case study illustrates how FAS can exploit its additional spatial DoFs together with signal-domain modulation to support joint communication and sensing. This perspective may inspire future frameworks using alternative modulation, waveforms, or learning-assisted approaches.
\vspace{-0.75em}
\section{Open Challenges and Future Directions}

Existing FAS-ISAC studies have mainly examined positional reconfiguration through snapshot-based optimization and simplified network settings. To advance toward predictive, coordinated, and secure spatial-state management, this section outlines three future directions: closed-loop active sensing, conflict-aware spatial-state allocation, and privacy of spatial-state adaptation.
\vspace{-1.5em}

\subsection{From Passive Estimation to Closed-Loop Active Sensing}

Temporal consistency is essential for translating the spatial agility of FAS into reliable ISAC gains. Existing FAS-ISAC designs largely rely on instantaneous optimization, where antenna positions, beamforming vectors, and sensing objectives are determined for a given system state. While analytically convenient, most existing formulations do not explicitly account for practical implementation overheads, such as position-dependent CSI acquisition, calibration, sensing latency, and FAS reconfiguration latency. Consequently, the selected spatial state may become outdated by the time it is executed, weakening predict-then-optimize frameworks and degrading sensing accuracy and communication reliability.

This temporal mismatch motivates a move from static spatial optimization toward predictive closed-loop control. Instead of seeking the optimal position for a current snapshot, future architectures can determine which antenna-positioning action maximizes the utility of subsequent observations and transmissions. FAS can thus extend sensing beyond passive estimation toward active information acquisition, where antenna positions reshape the geometric observation model. Future research could formalize this within belief-driven frameworks, where position sequences maximize future information gain while maintaining tracking stability. Such frameworks should jointly account for CSI acquisition, calibration, and reconfiguration overhead, while constraining spatial-state prediction by finite-aperture correlation, mutual coupling, and switching losses.
\vspace{-2.25em}
\subsection{Conflict-Aware Spatial Resource Allocation in Multi-User and Multi-Target Scenarios}
As FAS-ISAC evolves toward dense multi-user and multi-target deployments, spatial flexibility becomes a contested physical-layer resource rather than merely an additional optimization DoF. Existing studies often optimize an antenna position, selected port, or activated port set for a dominant communication or sensing objective. In practical ISAC networks, however, heterogeneous users, multiple targets, diverse quality-of-service requirements, and time-varying priorities may coexist. A spatial state benefiting one task may impair another; for instance, a port strengthening a communication link may not provide favorable sensing geometry, whereas a sensing-oriented state may weaken multi-user interference suppression. Thus, the challenge shifts from identifying the best position for an isolated task to allocating limited spatial resources among dual-functional tasks.

This issue distinguishes FAS-ISAC from multi-user or multi-target ISAC design, where conflicts mainly arise in beamforming, power allocation, and time-frequency resource sharing. By contrast, FAS-ISAC introduces a reconfigurable spatial domain, where ports, port sets, sub-apertures, and fluidic configurations serve as allocatable resources. Directions include weighted utility maximization, max-min fairness, worst-user or worst-target guarantees, multi-port activation, port clustering, hybrid fluid-fixed arrays, and multi-state fluidic configurations. Spatial scheduling and network-level geometry allocation may allow users, targets, or distributed ISAC nodes to share spatial flexibility over time and space, shifting the focus toward communication-sensing coordination.
\vspace{-1.5em}

\subsection{Security and Privacy in FAS Configuration Adaptation}
The positional flexibility of FAS reshapes the security landscape of ISAC systems. Configuration adaptation can improve secrecy, interference avoidance, and robustness against adverse propagation environments. At the same time, its controllability and observability may enlarge the attack surface of FAS-ISAC. Existing ISAC security studies mainly focus on signal transmission, sensing integrity, or communication privacy, whereas FAS-ISAC further exposes configuration adaptation to manipulation and inference. Spoofed echoes, falsified CSI, user impersonation, or malicious reflections may mislead the system toward adversarial spatial states, degrading communication and sensing performance.

Future FAS-ISAC systems should protect not only transmitted signals but also antenna-configuration adaptation. Important directions include robust spatial decision-making under deceptive observations, randomized spatial-state selection, privacy-preserving control signaling, and adversarially resilient learning. Port activation sequences or spatial-state transitions may also act as side channels, leaking information about users, targets, sensing priorities, or environmental geometry. Consequently, secure and privacy-preserving spatial-state management will become essential for FAS-ISAC networks.
\vspace{-0.75em}

\section{Conclusion}
FAS-ISAC reframes the spatial domain from an FPA-constrained resource into an adaptive design space for joint sensing and communication. Beyond increasing the accessible spatial states, FAS-ISAC introduces spatial reconfigurability as a design dimension for coordinating communication and sensing across physical-layer optimization, signal design, and system implementation. The FA-MM case study further illustrated how spatial states can be coupled with modulation for dual-functional transceiver design toward 6G and beyond.

\vspace{-1em}
\bibliography{Refer.bib}
\vspace{-1.5em}
\section*{Biography}
Yueling Zhao (202411220649@std.uestc.edu.cn) is pursuing the Ph.D. degree with the National Key Laboratory of Wireless Communications, University of Electronic Science and Technology of China (UESTC), Chengdu, China. His research interests include fluid antennas, ISAC, and waveform design.

Ping Yang (yang.ping@uestc.edu.cn) is a Professor with the National Key Laboratory of Wireless Communications, UESTC, and served as an Editor for IEEE Communications Letters and IEEE JSTSP. His research interests include 5G/6G systems, machine learning, and bionic communications.

Hong Niu (hong.niu@ntu.edu.sg) is a Research Fellow with the School of Electrical and Electronic Engineering, Nanyang Technological University, Singapore. His research interests include physical-layer security and quantum communications.

Shuaixin Yang (shuaixin.yang@foxmail.com) received the B.E. and Ph.D. degrees from UESTC in 2020 and 2026, respectively. His research interests include MIMO, emerging antennas, IRS, and wireless optimization.

Yue Xiao (xiaoyue@uestc.edu.cn) serves as a Professor with the National Key Laboratory of Wireless Communications, UESTC. His research interests include wireless system design and signal processing for future wireless communications.

Gang Wu (wugang99@uestc.edu.cn) holds a Professorship with the National Key Laboratory of Wireless Communications, UESTC. His research interests include PHY/MAC techniques, resource allocation, and 5G/6G wireless systems.

Mérouane Debbah (merouane.debbah@ku.ac.ae) is a Professor with Khalifa University, Abu Dhabi, UAE, and the Founding Senior Director of the KU Digital Future Institute. His research interests include 5G/6G, large intelligent surfaces, AI, and semantic communications.

Ertugrul Basar (ertugrul.basar@tuni.fi) is a Professor of Wireless Systems at Tampere University and the Head of the Wireless Systems Group, Tampere Wireless Research Center, Finland. His research interests include 6G, RIS, MIMO, quantum communications, and AI-enabled wireless networks.

\end{document}